\documentclass[twocolumn,showpacs,preprintnumbers]{revtex4}
\usepackage{amsmath,amssymb}
\usepackage{graphicx}
\usepackage{dcolumn}
\usepackage[dvips]{epsfig,color}
\usepackage{bm}
\begin{document}
\title{Relativistic quark-diquark model of baryons with a spin-isospin transition interaction}   
\author{M. De Sanctis}
\affiliation{Universidad Nacional de Colombia, Bogot\'a, Colombia}
\author{J. Ferretti}
\affiliation{Dipartimento di Fisica and INFN, Universit\'a La Sapienza, Piazzale A. Moro 5, 00185, Roma, Italy} 
\author{E. Santopinto}\thanks{Corresponding author: santopinto@ge.infn.it}\author{A. Vassallo}
\affiliation{INFN, Sezione di Genova, via Dodecaneso 33, 16146 Genova (Italy)}    
\begin{abstract}
The relativistic interacting quark-diquark model of baryons, recently developed, is here extended to introduce a spin-isospin 
transition interaction into the mass operator. The refined version of the model is used to calculate the non strange baryon spectrum. 
The results are compared to the present experimental data. 
\end{abstract}
\pacs{12.39.Ki, 12.39.Pn, 14.20.Gk}
\maketitle

\section{Introduction}
According to quark models (QM's) \cite{Isgur:1978xj,Capstick:1986bm,Bijker:1994yr,Ferraris:1995ui,Glozman:1995fu,Loring:2001kv}, 
baryons can be described as the bound states of three constituent quarks. These are effective degrees of freedom that mimic the three 
valence quarks inside baryons, with a sea of gluons and $q \bar q$ sea pairs.
The light baryons can then be ordered according to the approximate SU$_{\mbox{f}}$(3) symmetry into the multiplets 
$[1]_A \oplus [8]_M \oplus [8]_M \oplus [10]_S$. 
QM's explain quite well several properties of baryons, such as the strong decays and the magnetic moments.
Nevertheless, they predict a larger number of states than the experimentally observed ones (the missing resonances problem) and states with 
certain quantum numbers appear in the spectrum at excitation energies much lower than predicted \cite{Nakamura:2010zzi}. 
The problem of the missing resonances \cite{Nakamura:2010zzi,Capstick:1992uc,Capstick:1992th} has motivated the realization of 
several experiments, such as CB-ELSA \cite{Crede:2003ax}, TAPS \cite{Krusche:1995nv}, GRAAL \cite{Renard:2000iv}, SAPHIR \cite{Tran:1998qw} 
and CLAS \cite{Dugger:2002ft}, which only provided a few weak indications about some states. Indeed, even if several experiments have been 
dedicated to the search of missing states, just a small number of new resonances has been included into the PDG \cite{Nakamura:2010zzi}.

There are two possible explanations to the puzzle of the missing resonances: 1) There may be resonances very weakly coupled to the single 
pion, but with higher probabilities of decaying into two or more pions or into other mesons 
\cite{Nakamura:2010zzi,Capstick:1992uc,Capstick:1992th}. 
The detection of such states is further complicated by the problem of the separation of the experimental data from the background and by 
the expansion of the differential cross section into many partial waves; 2) Alternately, it is possible to consider models that are 
characterized by a smaller number of effective degrees of freedom with respect to the three quarks QM's and to assume that the 
majority of the missing states, not yet experimentally observed, simply may not exist. 
This is the case of quark-diquark models \cite{Ida:1966ev,Anselmino:1992vg,Ferretti:2011zz,DeSanctis:2011zz,
Galata:2012xt,Jaffe:2004ph,Wilczek:2004im,Selem:2006nd,Santopinto:2004hw,Cloet:2009,Forkel:2008un,Anisovich:2010wx}, 
where two quarks are strongly correlated and thus the state space is heavily reduced.

In quark-diquark models, the effective degrees of freedom of diquarks are introduced to describe baryons as bound states of a 
constituent diquark and quark \cite{Ida:1966ev}.
The notion of diquark dates back to 1964, when its possibility was mentioned by Gell-Mann \cite{GellMann:1964nj} in his original paper on 
quarks. 
Since then, many papers have been written on this topic (for a review see Ref. \cite{Anselmino:1992vg}) and, more recently, the diquark 
concept has been applied to various calculations 
\cite{Jakob:1997,Bloch:1999ke,Brodsky:2002,Ma,Oettel:2002wf,Gamberg:2003,Jaffe:2003sg,Wilczek:2004im,Maris:2004,Jaffe:2004ph,Selem:2006nd,
DeGrand:2007vu,BacchettaRadici,Forkel:2008un,Cloet:2009,Anisovich:2010wx,Santopinto:2004hw,Ferretti:2011zz,DeSanctis:2011zz,Galata:2012xt}. 
Important phenomenological indications for diquark-like correlations have been collected 
\cite{Jaffe:2004ph,Selem:2006nd,Close:1988br,Neubert} and indications for diquark confinement have also been provided 
\cite{Bender:1996bb}. This makes plausibly enough to make diquarks a part of the baryon's wave function.

In Ref. \cite{Santopinto:2004hw}, one of us developed an nonrelativistic interacting quark-diquark model, i.e. a potential model based on 
the effective degrees of freedom of a constituent quark and diquark. In Ref. \cite{Ferretti:2011zz}, it was "relativized" and 
reformulated within the point form formalism \cite{Klink:1998zz}. In Ref. \cite{DeSanctis:2011zz}, we used the wave functions of Ref. 
\cite{Ferretti:2011zz} to compute the nucleon electromagnetic form factors. 
Here, we intend to improve the "relativized" model \cite{Ferretti:2011zz,DeSanctis:2011zz} and compute the non strange baryon spectrum 
within point form dynamics.  

Even if our previous results for the non strange baryon spectrum \cite{Ferretti:2011zz} were in general quite good, here we intend to show 
how the introduction of a spin-isospin transition interaction, inducing the mixing between quark-scalar diquark and quark-axial-vector 
diquark states in the nucleon wave function, can further improve them, as already suggested in Ref. \cite{Santopinto:2004hw}. 
Scalar and axial-vector diquarks are two correlated quarks in $S$ wave with spin 0 or 1, respectively \cite{Wilczek:2004im,Jaffe:2004ph}.
In a following paper, we will use the new wave functions, obtained by solving the eigenvalue problem of the mass operator of the present 
model, to compute the nucleon electromagnetic form factors and the elicity amplitudes.

\section{The Mass operator}
\label{The Model} 
We consider a quark-diquark system, where $\vec{r}$ is the relative coordinate between the two constituents and $\vec{q}$ is the conjugate 
momentum to $\vec{r}$. 
We propose a relativistic quark-diquark model, based on the following baryon rest frame mass operator 
\begin{equation}
	\begin{array}{rcl}
	M & = & E_0 + \sqrt{\vec q\hspace{0.08cm}^2 + m_1^2} + \sqrt{\vec q\hspace{0.08cm}^2 + m_2^2} + M_{\mbox{dir}}(r)  \\
	  & + & M_{\mbox{cont}}(q,r) + M_{\mbox{ex}}(r) + M_{\mbox{tr}}(r)~,  
	\end{array}
	\label{eqn:H0}
\end{equation}
where $E_0$ is a constant, $M_{\mbox{dir}}(r)$ and $M_{\mbox{ex}}(r)$ respectively the direct and the exchange diquark-quark interaction, 
$m_1$ and $m_2$ stand for diquark and quark masses, where $m_1$ is either $m_S$ or $m_{AV}$ according if the part of the mass operator 
diagonal in the diquark spin [i.e. the whole mass operator of Eq. (\ref{eqn:H0}) without the interaction $M_{\mbox{tr}}(r)$] acts on a 
scalar or an axial-vector diquark 
\cite{Jaffe:2004ph,Lichtenberg:1979de,deCastro:1993sr,Schafer:1993ra,Cahill:1995ka,Lichtenberg:1996fi,Burden:1996nh,Maris,Orginos:2005vr,Wilczek:2004im,Flambaum:2005kc,Eichmann:2008ef,Babich:2007ah}, 
$M_{\mbox{cont}}(q,r)$ is a contact interaction and $M_{\mbox{tr}}(r)$ is a spin-isospin transition interaction.

The direct term is a Coulomb-like interaction with a cut off plus a linear confinement term
\begin{equation}
  \label{eq:Vdir}
  M_{\mbox{dir}}(r)=-\frac{\tau}{r} \left(1 - e^{-\mu r}\right)+ \beta r ~~.
\end{equation}

One needs also an exchange interaction \cite {Lichtenberg:1981pp,Santopinto:2004hw}, since this is the crucial ingredient of a 
quark-diquark description of baryons. We have
\begin{gather}
  M_{\mbox{ex}}(r) = (-1)^{l + 1} e^{-\sigma r} 
  \left [ A_S \mbox{ } \vec s_1 \cdot \vec s_2 + A_I \mbox{ } \vec t_1 \cdot \vec t_2 \right . 
    \notag \\ 
    +\left . A_{SI} \mbox{ } \vec s_1 \cdot \vec s_2 \mbox{ } \vec t_1 \cdot \vec t_2 \right ]  ~~,
    \label{eqn:VexchS1}
\end{gather}
where $\vec{s}$ and $\vec{t}$ are the spin and the isospin operators. 

Moreover, we consider a contact interaction similar to that introduced by Godfrey and Isgur \cite{Godfrey:1985xj} 
\begin{equation}
	\label{eqn:Vcont(r)}
	\begin{array}{rcl}
	M_{\mbox{cont}} & = & \left(\frac{m_1 m_2}{E_1 E_2}\right)^{1/2}  
	\frac{\eta^3 D}{\pi^{3/2}} e^{-\eta^2 r^2} \mbox{ } \delta_{L,0} \delta_{s_1,1} 
	\left(\frac{m_1 m_2}{E_1 E_2}\right)^{1/2} \mbox{ },
	\end{array}
\end{equation}
where $E_i = \sqrt{\vec q\hspace{0.08cm}^2 + m_i^2}$ ($i$ = 1, 2), $\eta$ and $D$ are parameters of the model.

Finally we consider a spin-isospin transition interaction, $M_{\mbox{tr}}(r)$, in order to mix quark-scalar diquark and quark-axial-vector 
diquark states. $M_{\mbox{tr}}(r)$ is chosen as
\begin{equation}    
	\label{eqn:Vtr(r)}
	M_{\mbox{tr}}(r) = V_0 \mbox{ } e^{-\frac{1}{2} \nu^2 r^2} (\vec s_2 \cdot \vec S) 
	(\vec t_2 \cdot \vec T) \mbox{ },
\end{equation}
where $V_0$ and $\nu$ are free parameters.
The matrix elements of the spin transition operator, $\vec S$, are defined as:
\begin{subequations}
\begin{equation}
	\left\langle \right. s_1', m_{s_1}' \left. \right| S_\mu^{[1]} \left| s_1, m_{s_1} \right\rangle \neq 0 
	\mbox{ for } s_1' \neq s_1  \mbox{ },
\end{equation}
where
\begin{equation}
	\left\langle 1 \right\| S_1 \left\| 0 \right\rangle = 1  \mbox{ },
\end{equation}
\begin{equation}
	\left\langle 0 \right\| S_1 \left\| 1 \right\rangle = -1  
\end{equation}
\end{subequations}
and the same holds for those of the isospin transition operator, $\vec T$. Thus one has:
\begin{equation}
	\begin{array}{rcl}
	\left\langle \Phi' \right| M_{\mbox{tr}} \left| \Phi \right\rangle & = & \frac{1}{4} V_0 \delta_{s_1',s_1\pm1} 
	\delta_{S\frac{1}{2}} \delta_{t_1',t_1\pm1} \delta_{T\frac{1}{2}} \\ 
	& \times & \left\langle \Phi'(\vec r) \right| e^{-\frac{1}{2} \nu^2 r^2} 
	\left| \Phi(\vec r) \right\rangle  \mbox{ },
	\end{array}
\end{equation}
where $\Phi(\vec r)$ stands for the spatial wave function of the generic state, $\left| \Phi \right\rangle$. 

\begin{figure}[htbp] 
\centering 
\includegraphics[width=7cm]{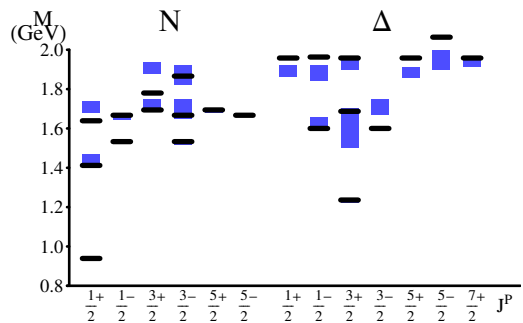}
\caption{(Color online) Comparison between the calculated masses (black lines) of the $3^*$ and $4^*$ non strange baryon resonances (up to 
2 GeV) and the experimental masses from PDG \cite{Nakamura:2010zzi} (boxes).} 
\label{fig:Spectrum3e4}
\end{figure}

The mass formula of the previous version of the relativistic quark-diquark model \cite{Ferretti:2011zz} is
\begin{widetext}
\begin{equation}
	\label{eqn:RMF}
	\begin{array}{rcl}
	M & = & E_0 + \sqrt{\vec q\hspace{0.08cm}^2 + m_1^2} + \sqrt{\vec q\hspace{0.08cm}^2 + m_2^2} 
	- \frac{\tau}{r} \left(1 - e^{-\mu r}\right) 
	+ \beta r + \left(\frac{m_1 m_2}{E_1 E_2}\right)^{1/2+\epsilon}  
	\frac{\eta^3 D}{\pi^{3/2}} e^{-\eta^2 r^2} \mbox{ } \delta_{L,0} \delta_{s_1,1} 
	\left(\frac{m_1 m_2}{E_1 E_2}\right)^{1/2+\epsilon} \\
	& + & (-1)^{l + 1} e^{-\sigma r} \left [ A_S \mbox{ } \vec s_1 \cdot \vec s_2 + A_I \mbox{ } \vec t_1 \cdot \vec t_2
	+ A_{SI} \mbox{ } \vec s_1 \cdot \vec s_2 \mbox{ } \vec t_1 \cdot \vec t_2 \right ] ~~.
	\end{array}
\end{equation}
\end{widetext}
The main difference between the mass operator of Eq. (\ref{eqn:H0}) and that of Eq. (\ref{eqn:RMF}) \cite{Ferretti:2011zz} is the presence 
of the spin-isospin transition interaction $M_{\mbox{tr}}$ in Eq. (\ref{eqn:H0}). 
$M_{\mbox{tr}}(r)$ is introduced to improve the description of the electromagnetic elastic form factors of the nucleon 
\cite{DeSanctis:2011zz,DeSanctis:TBP}. Indeed, $M_{\mbox{tr}}(r)$ makes it possible to have a nucleon wave function with a 
quark-axial-vector diquark component in addition to the quark-scalar diquark one. 
At the same time, $M_{\mbox{tr}}(r)$ significantly improves the description of the non strange baryon spectrum \cite{Ferretti:2011zz} 
(see Fig. \ref{fig:Spectrum3e4}).

One can also notice that the values of the model parameters change significantly from those of Ref. \cite{DeSanctis:2011zz,Ferretti:2011zz} 
after the introduction of the interaction (\ref{eqn:Vtr(r)}) into the mass formula. 
In particular, one can see that the masses of the two constituents (the quark and the diquark) are now smaller than before, which is good 
in a relativistic QM, and the mass difference between the scalar and the axial-vector diquark is smaller too (it goes from 350 MeV to 210 
MeV). The same happens for the string tension, that goes from 2.15 $\mbox{fm}^{-2}$ to 1.57 $\mbox{fm}^{-2}$. 

It is worth noting that the number of model parameters increases only by one, since there are two new parameters, $V_0$ and $\nu$ 
[see Eq. (\ref{eqn:Vtr(r)})], while the parameter $\epsilon$ of the contact interaction [see Eqs. (\ref{eqn:Vcont(r)}) and (\ref{eqn:RMF})] 
has been removed. 
\begin{table}[h]  
\begin{center}
\begin{tabular}{llllll}
\hline
\hline \\
$m_q$ & $=140$ MeV & $~m_{S}$ & $=150$ MeV & $~m_{AV}$ & $=360$ MeV \\ 
$~\tau$ & $=1.23$   & $~\mu$ & $=125~\mbox{fm}^{-1}$ & $~\beta$ & $=1.57~\mbox{fm}^{-2}$ \\ 
$A_S$ & $=125$ MeV & $A_I$ & $=85$ MeV & $A_{SI}$ & $=350$ MeV \\
$~\sigma$ & $=0.60~\mbox{fm}^{-1}$ & $~E_0$ & $=826$ MeV & $D$ & $=2.00$ $\mbox{fm}^2$ \\ 
$~\eta$ & $=10.0~\mbox{fm}^{-1}$ & $~V_0$ & $=1450$ MeV  & $\nu$ & $=0.35$ $\mbox{fm}^{-1}$ \\ \\
\hline
\hline
\end{tabular}
\end{center}
\caption{Resulting values for the model parameters.}
\label{tab:ResultingParameters}
\end{table}
Finally, it has to be noted that in the present work all the calculations are performed without any perturbative approximation, as in Ref. 
\cite{Ferretti:2011zz}. 

The eigenfunctions of the mass operator of Eq. (\ref{eqn:H0}) can be thought as eigenstates of the mass operator with interaction in a 
Bakamjian-Thomas construction \cite{BT}. The interaction is introduced adding an interaction term to the free mass operator 
$M_0 = \sqrt{\vec q\hspace{0.08cm}^2 + m_1^2} + \sqrt{\vec q\hspace{0.08cm}^2 + m_2^2}$, in such a way that the interaction commutes with 
the non interacting Lorenz generators and with the non interacting four velocity \cite{KP}.

The dynamics is given by a point form Bakamjian-Thomas construction. Point form means that the Lorentz group is kinematic. 
Furthermore, since we are doing a point form Bakamjian-Thomas construction, here $P = M V_0$ where $V_0$ is the noninteracting 
four-velocity (whose eigenvalue is $v$).

The general quark-diquark state, defined on the product space $H_1 \otimes H_2$ of the one-particle spin $s_1$ (0 or 1) and spin $s_2$ 
(1/2) positive energy representations $H_1=L^2(R^3)\otimes S_1^{0}$ or $H_1=L^2(R^3)\otimes S_1^{1}$ and $H_2=L^2(R^3)\otimes S_2^{1/2}$ of 
the Poincar\'e Group, can be written as \cite{Ferretti:2011zz}
\begin{equation}
	\left|  p_1, p_2, \lambda_1, \lambda_2 \right\rangle \mbox{ },
\end{equation} 
where $p_1$ and $p_2$ are the four-momenta of the diquark and the quark, respectively, while $\lambda_1$ and $\lambda_2$ are, respectively, the $z$-projections of their spins.

We introduce the velocity states as \cite{Klink:1998zz,Ferretti:2011zz} 
\begin{equation}
	\label{eqn:velocity-states}
	\vert v,\vec{k}_1,\lambda_1,\vec{k}_2,\lambda_2\rangle =
	U_{B(v)}\vert k_1,s_1,\lambda_1, k_2,s_2,\lambda_2 \rangle_{0}  \mbox{ },
\end{equation} 
where the suffix $0$ means that the diquark and the quark three-momenta $\vec {k}_1$ and $\vec{k}_2$, called internal momenta, satisfy: 
\begin{equation}
	\vec {k}_1 + \vec{k}_2=0 \mbox{ }.
\end{equation}
Following the standard rules of the point form approach, the boost operator $U_{B(v)}$ is taken as a canonical one, obtaining that the transformed four-momenta are given by $p_{1,2}=B(v)k_{1,2}$ and satisfy the point form relation
\begin{equation}
	\label{eq:pfe}
	p_1^\mu + p_2^\mu = \frac{P_N^\mu}{M_N} \left( \sqrt{\vec q\hspace{0.08cm}^2 + m_1^2} + \sqrt{\vec q\hspace{0.08cm}^2 + m_2^2} \right)  \mbox{ },
\end{equation}
where $P^\mu_N$ is the observed nucleon four-momentum and $M_N$ is its mass. 
It is worthwhile noting that Eq. (\ref{eqn:velocity-states}) redefines the single particle spins. 
Having applied canonical boosts, the conditions for a point form approach \cite{Klink:1998zz,melde} are satisfied.
Therefore, the spins on the left hand state of Eq. (\ref{eqn:velocity-states}) perform the same Wigner rotations as $\vec k_1$ and $\vec k_2$, allowing to couple the spin and the orbital angular momentum as in the non relativistic case \cite{Klink:1998zz}, while the spins in the ket on the right hand of Eq. (\ref{eqn:velocity-states}) undergo the single particle Wigner rotations.

In Point form dynamics, Eq. (\ref{eqn:H0}) corresponds to a good mass operator since it commutes with the Lorentz generators and with the 
four velocity. 
We diagonalize Eq. (\ref{eqn:H0}) in the Hilbert space spanned by the velocity states. 
Finally, instead of the internal momenta $\vec{k_1}$ and $\vec{k_2}$ we use the relative momentum $\vec q$, conjugate to the relative 
coordinate $\vec{r} = \vec{r}_1 - \vec{r}_2$, thus considering the following velocity basis states:
\begin{equation}
	\vert v,\vec q,\lambda_1,\lambda_2 \rangle =
	U_{B(v)} \vert k_1,s_1,\lambda_1,k_2,s_2,\lambda_2 \rangle_{0} \mbox{ }.
\end{equation} 

\section{Results and discussion}
Figure \ref{fig:Spectrum3e4} and Table \ref{tab:Spectrum} show the comparison between the experimental data \cite{Nakamura:2010zzi,Anisovich:2011fc} 
and the results of our quark-diquark model calculation, obtained with the set of parameters of Table \ref{tab:ResultingParameters}. 
In addition to the experimental data from PDG \cite{Nakamura:2010zzi}, we also consider the latest multi-channel Bonn-Gatchina partial wave 
analysis results, including data from Crystal Barrel/TAPS at ELSA and other laboratories \cite{Anisovich:2011fc}. 
In particular, these data differ from those of the PDG \cite{Nakamura:2010zzi} in the case of the $\Delta(1940) D_{33}$.

\begin{table}[h1]
\begin{center}
\begin{tabular}{ccccccccc}
\hline
\hline \\
Resonance & Status & $M^{\mbox{exp.}}$ & $J^P$ & $L^P$ & $S$ & $s_1$ & $n_r$ & $M^{\mbox{calc.}}$ \\
 &  & (MeV) &  &  &  &  &  & (MeV) \\ \\
\hline \\
$N(939)$       $P_{11}$ & **** &  939         &  $\frac{1}{2}^+$ & $0^+$ &
$\frac{1}{2}$  & 0,1 & 0  & 939  \\
$N(1440)$      $P_{11}$ & **** &  1420 - 1470 &  $\frac{1}{2}^+$ & $0^+$ &
$\frac{1}{2}$  & 0,1 & 1  & 1412  \\
$N(1520)$      $D_{13}$ & **** &  1515 - 1525 &  $\frac{3}{2}^-$ & $1^-$ & 
$\frac{1}{2}$  & 0,1 & 0  & 1533  \\
$N(1535)$      $S_{11}$ & **** &  1525 - 1545 &  $\frac{1}{2}^-$ & $1^-$ & 
$\frac{1}{2}$  & 0,1 & 0  & 1533  \\
$N(1650)$      $S_{11}$ & **** &  1645 - 1670 &  $\frac{1}{2}^-$ & $1^-$ & 
$\frac{3}{2}$  & 1 & 0  & 1667  \\
$N(1675)$      $D_{15}$ & **** &  1670 - 1680 &  $\frac{5}{2}^-$ & $1^-$ & 
$\frac{3}{2}$  & 1 & 0  & 1667  \\
$N(1680)$      $F_{15}$ & **** &  1680 - 1690 &  $\frac{5}{2}^+$ & $2^+$ & 
$\frac{1}{2}$  & 0,1 & 0  & 1694  \\
$N(1700)$      $D_{13}$ & ***  &  1650 - 1750 &  $\frac{3}{2}^-$ & $1^-$ & 
$\frac{3}{2}$  & 1 & 0  & 1667  \\
$N(1710)$      $P_{11}$ & ***  &  1680 - 1740 &  $\frac{1}{2}^+$ & $0^+$ & 
$\frac{1}{2}$  & 0,1 & 2  & 1639  \\
$N(1720)$      $P_{13}$ & **** &  1700 - 1750 &  $\frac{3}{2}^+$ & $2^+$ & 
$\frac{1}{2}$  & 0,1 & 0  & 1694  \\
$N(1875)$      $D_{13}$ & ***  &  1820 - 1920 &  $\frac{3}{2}^-$ & $1^-$ & 
$\frac{1}{2}$  & 0,1 & 1  & 1866  \\
$N(1880)$      $P_{11}$ & **   & 1835 - 1905  &  $\frac{1}{2}^+$ & $0^+$ &
$\frac{1}{2}$  & 0,1 & 3  &  1786  \\
$N(1895)$      $S_{11}$ & **    & 1880 - 1910  &  $\frac{1}{2}^-$ & $1^-$ & 
$\frac{1}{2}$  & 0,1 & 1  & 1866  \\
$N(1900)$ $P_{13}$   & *** &   1875 - 1935    &  $\frac{3}{2}^+$ & $0^+$ &
$\frac{3}{2}$  & 0 & 0  &  1780  \\
missing        &  -- & -- &  $\frac{3}{2}^+$ & $2^+$ & $\frac{1}{2}$  & 0,1 & 1  &  1990  \\
$N(2000)$      $F_{15}$ & **   &  1950 - 2150 &  $\frac{5}{2}^+$ & $2^+$ & 
$\frac{1}{2}$  & 0,1 & 1  & 1990  \\  \\

$\Delta(1232)$ $P_{33}$ & **** &  1230 - 1234 &  $\frac{3}{2}^+$ & $0^+$ & 
$\frac{3}{2}$  & 1 & 0  & 1236  \\
$\Delta(1600)$ $P_{33}$ & ***  &  1500 - 1700 &  $\frac{3}{2}^+$ & $0^+$ & 
$\frac{3}{2}$  & 1 & 1  & 1687  \\
$\Delta(1620)$ $S_{31}$ & **** &  1600 - 1660 &  $\frac{1}{2}^-$ & $1^-$ & 
$\frac{1}{2}$  & 1 & 0  & 1600 \\
$\Delta(1700)$ $D_{33}$ & **** &  1670 - 1750 &  $\frac{3}{2}^-$ & $1^-$ & 
$\frac{1}{2}$  & 1 & 0  & 1600 \\
$\Delta(1750)$ $P_{31}$ & *  & 1708 - 1780  &  $\frac{1}{2}^+$ & $0^+$ & 
$\frac{1}{2}$  & 1 & 0  & 1857  \\
$\Delta(1900)$ $S_{31}$ & **   &  1840 - 1920 &  $\frac{1}{2}^-$ & $1^-$ & 
$\frac{1}{2}$  & 1 & 1  & 1963 \\
$\Delta(1905)$ $F_{35}$ & **** &  1855 - 1910 &  $\frac{5}{2}^+$ & $2^+$ & 
$\frac{3}{2}$  & 1 & 0  & 1958  \\
$\Delta(1910)$ $P_{31}$ & **** &  1860 - 1920 &  $\frac{1}{2}^+$ & $2^+$ & 
$\frac{3}{2}$  & 1 & 0  & 1958  \\
$\Delta(1920)$ $P_{33}$ & ***  &  1900 - 1970 &  $\frac{3}{2}^+$ & $2^+$ & 
$\frac{3}{2}$  & 1 & 0  & 1958  \\
$\Delta(1930)$ $D_{35}$ & ***  &  1900 - 2000 &  $\frac{5}{2}^-$ & $1^-$ & 
$\frac{3}{2}$  & 1 & 0  & 2064  \\
$\Delta(1940)$ $D_{33}$ & **  &  1940 - 2060  &  $\frac{3}{2}^-$ & $1^-$ & 
$\frac{1}{2}$  & 1 & 1  & 1963 \\ 
$\Delta(1950)$ $F_{37}$ & **** &  1915 - 1950 &  $\frac{7}{2}^+$ & $2^+$ & 
$\frac{3}{2}$  & 1 & 0  & 1958  \\  \\
\hline
\hline
\end{tabular}
\end{center}
\caption{Comparison between the experimental \cite{Nakamura:2010zzi} values of non strange baryon resonances masses (up to 2 GeV) and the numerical ones (all values are expressed in $MeV$). Tentative assignments of $2^*$ and $1^*$ resonances are shown in the second part of the table. $J^P$ and $L^P$ are respectively the total angular momentum and the orbital angular momentum of the baryon, including the parity $P$; $S$ is the total spin, obtained coupling the spin of the diquark $s_1$ and that of the quark; finally $n_r$ is the number of nodes in the radial wave function.} 
\label{tab:Spectrum}
\end{table}

\begin{table}
\begin{center}
\begin{tabular}{ccc}
\hline
\hline
 &  &  \\
$m_S$ $\mbox{(MeV)}~~$ & $m_{AV} - m_S$ $\mbox{(MeV)}~~$ & Source \\ 
 &  &  \\
\hline \\
730   & 210 & Bloch {\it et al.} \cite{Bloch:1999ke} \\
750$\div$860 & 10$\div$170 & Oettel {\it et al.} \cite{Oettel:2002wf} \\
-     & 290 & Wilczek \cite{Wilczek:2004im} \\ 
-     & 210 & Jaffe \cite{Jaffe:2004ph} \\
600   & 350 & Ferretti {\it et al.} \cite{Ferretti:2011zz} \\  
852   & 224 & Galata and Santopinto \cite{Galata:2012xt} \\
-     & 200$\div$300 & Lichtenberg {\it et al.} \cite{Lichtenberg:1979de} \\
770   & 140 & de Castro {\it et al.} \cite{deCastro:1993sr} \\
420   & 520 & Sch\"{a}fer {\it et al.} \cite{Schafer:1993ra} \\
692   & 330 & Cahill {\it et al.} \cite{Cahill:1995ka} \\
595   & 205 & Lichtenberg {\it et al.} \cite{Lichtenberg:1996fi} \\
737   & 212 & Burden {\it et al.} \cite{Burden:1996nh} \\
688   & 202 & Maris \cite{Maris} \\
-     & 360 & Orginos \cite{Orginos:2005vr} \\
750   & 100 & Flambaum {\it et al.} \cite{Flambaum:2005kc} \\
590   & 210 &  \\
-     & 162 & Babich {\it et al.} \cite{Babich:2007ah} \\
-     & 270 & Eichmann {\it et al.} \cite{Eichmann:2008ef} \\
740   & 210 & Hecht {\it et al.} \cite{Hecht:2002ej} \\
-     & 135 & Santopinto and Galata \cite{Santopinto:2011mk} \\
710   & 199 & Ebert {\it et al.} \cite{Ebert} \\  
 --   & 183 & Chakrabarti {\it et al.} \cite{Chakrabarti:2010zz} \\
780   & 280 & Roberts {\it et al.} \cite{Roberts:2011cf} \\
150   & 210 & This work   \\  \\   
\hline
\hline
\end{tabular}
\end{center}
\caption{Mass difference (in MeV) between scalar and axial-vector diquarks, according to some previous studies.}
\label{tab:massediquarkaltri}
\end{table}

The spin-isospin transition interaction of Eq. (\ref{eqn:Vtr(r)}) mixes quark-scalar diquark and quark-axial-vector diquark states, i.e. 
states with $s_1 = 0$ ($t_1 = 0$) and $s_1 = 1$ ($t_1 = 1$), whose total spin (isospin) is $S = \frac{1}{2}$ ($T = \frac{1}{2}$). 
Thus, in this version of the model the nucleon state, as well as states such as the $D_{13}(1520)$, the $S_{11}(1535)$ and the 
$P_{11}(1440)$, contains both a $s_1 = 0$ and a $s_1 = 1$ component. 
In particular, the nucleon state, obtained by solving the eigenvalue problem of Eq. (\ref{eqn:H0}), in a schematic notation can be written 
as:
\begin{equation}
\label{eqn:nucleon.state}
	\left| N \right\rangle = 0.727 \left| qD_S, L=0 \right\rangle + 0.687 \left| qD_{AV}, L=0 \right\rangle  
	\mbox{ },
\end{equation}
where $D_S$ and $D_{AV}$ stand for the scalar and axial-vector diquarks, respectively, and $q$ for the quark.
The radial wave functions (in momentum space) of the quark-scalar diquark [$\Phi_S(q)$] and quark-axial-vector diquark [$\Phi_{AV}(q)$] 
systems of Eq. (\ref{eqn:nucleon.state}), obtained by solving the eigenvalue problem of Eq. (\ref{eqn:H0}), can be fitted by harmonic 
oscillator wave functions
\begin{subequations}
\begin{equation}
	\Phi_S(q) = \frac{2 \alpha_S^{3/2}}{\pi^{1/4}} \mbox{ } e^{-\frac{1}{2} \alpha_S^2 q^2}  \mbox{ },
\end{equation}
\begin{equation}
	\Phi_{AV}(q) = \frac{2 \alpha_{AV}^{3/2}}{\pi^{1/4}} \mbox{ } e^{-\frac{1}{2} \alpha_{AV}^2 q^2}  \mbox{ },
\end{equation}
\end{subequations} 
with $\alpha_S = 3.29$ GeV$^{-1}$ and $\alpha_{AV} = 3.04$ GeV$^{-1}$. This parametrization can then be used to compute observables, such 
as the nucleon electromagnetic form factors.  

The introduction of the interaction of Eq. (\ref{eqn:Vtr(r)}) determines an improvement in the overall quality of the reproduction of the 
experimental data (considering only $3^*$ and $4^*$ resonances), with respect to that obtained with the previous version of this model 
\cite{Ferretti:2011zz}. 
In particular, the Roper resonance, $N(1440)$ $P_{11}$, is far better reproduced than before and the same holds for $N(1680)$ $F_{15}$.

The present version of the relativistic quark-diquark model predicts only one missing state below the energy of 2 GeV 
(see Tab. \ref{tab:Spectrum}), while three quarks QM's give rise to several missing states \cite{Nakamura:2010zzi}. For example, Capstick 
and Isgur's model \cite{Capstick:1986bm} has 5 missing states up to 2 GeV, the hypercentral QM \cite{Giannini:2001kb} has 8, Glozman and 
Riska's model has 4 \cite{Glozman-Riska} and the U(7) model has 17 \cite{Bijker:1994yr}. The only missing resonance of our model, 
$N \frac{3}{2}^+(1990)$, lies at the same energy of the three star state $N(2000)$ $F_{15}$, which was previously a two star state of the 
PDG \cite{Nakamura:2010zzi}.
Indeed the two resonances, $N \frac{3}{2}^+(1990)$ and $N(2000)$ $F_{15}$, have the same quantum numbers, except for the total angular 
momentum, because their spin ($\frac{1}{2}$) and orbital angular momentum (2) are coupled to $J^P = \frac{3}{2}^+$ or $\frac{5}{2}^+$. 
Thus, to split the two resonances one should take a spin-orbit interaction into account.

While the absolute values of the diquark masses are model dependent, their difference is not. Comparing our result for the mass difference 
$m_{AV} - m_S$ between the axial-vector and the scalar diquark to those reported in Tab. \ref{tab:massediquarkaltri}, it is interesting to 
note that our estimation is comparable with all the others. Such evaluations come from phenomenological observations 
\cite{Jaffe:2004ph,Wilczek:2004im,Lichtenberg:1996fi}, lattice QCD calculations \cite{Orginos:2005vr,Babich:2007ah}, instanton liquid model 
calculations \cite{Schafer:1993ra}, applications of Dyson-Schwinger, Bethe-Salpeter and Fadde'ev equations 
\cite{Burden:1996nh,Hecht:2002ej,Maris,Cahill:1995ka,Flambaum:2005kc,Bloch:1999ke,Eichmann:2008ef} and constituent quark-diquark model 
calculations \cite{Ferretti:2011zz,Lichtenberg:1979de,deCastro:1993sr,Santopinto:2011mk}.

The whole mass operator of Eq. (\ref{eqn:H0}) is diagonalized by means of a numerical variational procedure, based on harmonic oscillator 
trial wave functions. With a variational basis made of $N = 200$ harmonic oscillator shells, the results converge very well.  



We think that the present paper can be helpful to the experimentalists in their analysis of the properties of the $N$ and $\Delta$-type 
resonances. Our quark-diquark model results may be compared to those of three quarks QM's, showing a larger number of missing resonances. 
Our results may then help the experimentalists to distinguish between the two interpretations for baryons.
Finally, in the future we will use our quark-diquark model wave functions to compute the nucleon electromagnetic form factors and the 
helicity amplitudes of baryon resonances \cite{DeSanctis:TBP}.




\begin{thebibliography}{50} 

\bibitem{Isgur:1978xj} 
  N.~Isgur and G.~Karl,
  Phys.\ Rev.\ D {\bf 18}, 4187 (1978);
  {\bf 19}, 2653 (1979);
  {\bf 20}, 1191 (1979).
	
\bibitem{Capstick:1986bm}
  S.~Capstick and N.~Isgur,
  Phys.\ Rev.\ D {\bf 34}, 2809 (1986).
	
\bibitem{Bijker:1994yr}   
  R.~Bijker, F.~Iachello and A.~Leviatan,
  Annals Phys.\  {\bf 236}, 69 (1994);	
  {\bf 284}, 89 (2000).
	
\bibitem{Ferraris:1995ui} 
  M.~Ferraris, M.~M.~Giannini, M.~Pizzo, E.~Santopinto and L.~Tiator,
  Phys.\ Lett.\ B {\bf 364}, 231 (1995);
  M.~M.~Giannini, E.~Santopinto and A.~Vassallo,
  Eur.\ Phys.\ J.\ A {\bf 12}, 447 (2001);     
  E.~Santopinto, A.~Vassallo, M.~M.~Giannini and M.~De Sanctis,
  Phys.\ Rev.\ C {\bf 76}, 062201 (2007);
  Phys.\ Rev.\ C {\bf 82}, 065204 (2010) .  
		
\bibitem{Glozman:1995fu} 
  L.~Y.~.Glozman and D.~O.~Riska,
  Phys.\ Rept.\  {\bf 268}, 263 (1996);
  L.~Y.~.Glozman, W.~Plessas, K.~Varga and R.~F.~Wagenbrunn,
  Phys.\ Rev.\ D {\bf 58}, 094030 (1998);
  R.~F.~Wagenbrunn, S.~Boffi, W.~Klink, W.~Plessas and M.~Radici,
  Phys.\ Lett.\  B {\bf 511}, 33 (2001);
  S.~Boffi, L.~Y.~Glozman, W.~Klink, W.~Plessas, M.~Radici and R.~F.~Wagenbrunn,
  Eur.\ Phys.\ J.\  A {\bf 14}, 17 (2002).		
					
\bibitem{Loring:2001kv} 
  U.~Loring, K.~Kretzschmar, B.~C.~Metsch and H.~R.~Petry,
  Eur.\ Phys.\ J.\ A {\bf 10}, 309 (2001);
  {\bf 10}, 395 (2001).

\bibitem{Nakamura:2010zzi}
  J.~Beringer {\it et al.} [Particle Data Group Collaboration],
  Phys.\ Rev.\ D {\bf 86}, 010001 (2012). 
	
\bibitem{Capstick:1992uc} 
  S.~Capstick,
  Phys.\ Rev.\ D {\bf 46}, 2864 (1992).
		
\bibitem{Capstick:1992th} 
  S.~Capstick and W.~Roberts,
  Phys.\ Rev.\ D {\bf 47}, 1994 (1993);
  {\bf 49}, 4570 (1994);
  {\bf 58}, 074011 (1998).
  
\bibitem{Crede:2003ax} 
  V.~Crede {\it et al.}  [CB-ELSA Collaboration],
  Phys.\ Rev.\ Lett.\  {\bf 94}, 012004 (2005);
  D.~Trnka {\it et al.}  [CBELSA/TAPS Collaboration],
  Phys.\ Rev.\ Lett.\  {\bf 94}, 192303 (2005).		
	
\bibitem{Krusche:1995nv} 
  B.~Krusche {\it et al.},
  Phys.\ Rev.\ Lett.\  {\bf 74}, 3736 (1995);
  F.~Harter {\it et al.},
  Phys.\ Lett.\ B {\bf 401}, 229 (1997);
  M.~Wolf {\it et al.},
  Eur.\ Phys.\ J.\ A {\bf 9}, 5 (2000).		
			
\bibitem{Renard:2000iv} 
  F.~Renard {\it et al.}  [GRAAL Collaboration],
  Phys.\ Lett.\ B {\bf 528}, 215 (2002);
  Y.~Assafiri {\it et al.},
  Phys.\ Rev.\ Lett.\  {\bf 90}, 222001 (2003).			
			
\bibitem{Tran:1998qw} 
  M.~Q.~Tran {\it et al.}  [SAPHIR Collaboration],
  Phys.\ Lett.\ B {\bf 445}, 20 (1998);
  K.~H.~Glander {\it et al.},
  Eur.\ Phys.\ J.\ A {\bf 19}, 251 (2004).  				
	
\bibitem{Dugger:2002ft} 
  M.~Dugger {\it et al.}  [CLAS Collaboration],
  Phys.\ Rev.\ Lett.\  {\bf 89}, 222002 (2002);
  {\bf 96}, 062001 (2006);
	M.~Ripani {\it et al.}  [CLAS Collaboration],
  Phys.\ Rev.\ Lett.\  {\bf 91}, 022002 (2003).  
  
\bibitem{Ida:1966ev}
  M.~Ida and R.~Kobayashi,
  Prog.\ Theor.\ Phys.\  {\bf 36}, 846 (1966);
  D. B. Lichtenberg and L. J. Tassie, 
  Phys. Rev. {\bf 155}, 1601 (1967).  
	
\bibitem{Anselmino:1992vg}
  M.~Anselmino, E.~Predazzi, S.~Ekelin, S.~Fredriksson and D.~B.~Lichtenberg,
  Rev.\ Mod.\ Phys.\  {\bf 65}, 1199 (1993).	
  
\bibitem{Jaffe:2004ph}  
  R.~L.~Jaffe,
  Phys.\ Rept.\ {\bf 409}, 1 (2005) 
  [Nucl.\ Phys.\ Proc.\ Suppl.\  {\bf 142}, 343 (2005)].
  
\bibitem{Wilczek:2004im}  
  F.~Wilczek,
  In *Shifman, M. (ed.) $et$ $al.$: From fields to strings, vol. 1* 77-93.  
  
\bibitem{Selem:2006nd}
  A.~Selem and F.~Wilczek,
  Ringberg 2005, New trends in HERA physics, pp. 337-356.  
   
\bibitem{Santopinto:2004hw}
  E.~Santopinto,
  Phys.\ Rev.\  C {\bf 72}, 022201 (2005).
  
\bibitem{Forkel:2008un}
  H.~Forkel and E.~Klempt,
  Phys.\ Lett.\  B {\bf 679}, 77 (2009).  
  
\bibitem{Cloet:2009}
  I.~C.~Cloet, G.~Eichmann, B.~El-Bennich, T.~Klahn and C.~D.~Roberts,
  Few Body Syst.\  {\bf 46}, 1 (2009).  

\bibitem{Anisovich:2010wx}
  A.~V.~Anisovich, V.~V.~Anisovich, M.~A.~Matveev, V.~A.~Nikonov, A.~V.~Sarantsev and 
  T.~O.~Vulfs,
  Int.\ J.\ Mod.\ Phys.\  A {\bf 25}, 2965 (2010) 
  [Int.\ J.\ Mod.\ Phys.\  A {\bf 25}, 3155 (2010)].
  
\bibitem{Ferretti:2011zz}   
  J.~Ferretti, A.~Vassallo and E.~Santopinto,
  Phys.\ Rev.\  C {\bf 83}, 065204 (2011);
  J.~Ferretti,
  Int.\ J.\ Mod.\ Phys.\ Conf.\ Ser.\  {\bf 26}, 1460117 (2014).

\bibitem{DeSanctis:2011zz}
  M.~De~Sanctis, J.~Ferretti, E.~Santopinto and A.~Vassallo,
  Phys.\ Rev.\  C {\bf 84}, 055201 (2011).  
  
\bibitem{Galata:2012xt} 
  G.~Galata and E.~Santopinto,
  Phys.\ Rev.\ C {\bf 86}, 045202 (2012). 
  
\bibitem{GellMann:1964nj}
  M.~Gell-Mann,
  Phys.\ Lett.\  {\bf 8}, 214 (1964).
  
\bibitem{Jakob:1997}
	R.~Jakob, P.~J.~Mulders and J.~Rodrigues,
  Nucl.\ Phys.\  A {\bf 626}, 937 (1997).
  
\bibitem{Bloch:1999ke}
  J.~C.~R.~Bloch, C.~D.~Roberts, S.~M.~Schmidt, A.~Bender and M.~R.~Frank,
  Phys.\ Rev.\  C {\bf 60}, 062201 (1999).  
  
\bibitem{Brodsky:2002}
	S.~J.~Brodsky, D.~S.~Hwang and I.~Schmidt,
  Phys.\ Lett.\  B {\bf 530}, 99 (2002).  
  
\bibitem{Ma}
	B.~Q.~Ma, D.~Qing and I. Schmidt, 
	Phys. Rev. C {\bf 65}, 035205 (2002).    
    
\bibitem{Oettel:2002wf}
  M.~Oettel and R.~Alkofer,
  Eur.\ Phys.\ J.\  A {\bf 16}, 95 (2003).  
  
\bibitem{Gamberg:2003}
  L.~P.~Gamberg, G.~R.~Goldstein and K.~A.~Oganessyan,
  Phys.\ Rev.\  D {\bf 67}, 071504 (2003). 
  
\bibitem{Jaffe:2003sg} 
  R.~L.~Jaffe and F.~Wilczek,
  Phys.\ Rev.\ Lett.\  {\bf 91}, 232003 (2003).  
  
\bibitem{Maris:2004}
  P.~Maris,
  Few Body Syst.\  {\bf 35}, 117 (2004).  
   
\bibitem{DeGrand:2007vu}
  T.~DeGrand, Z.~Liu and S.~Schaefer,
  Phys.\ Rev.\  D {\bf 77}, 034505 (2008).
  
\bibitem{BacchettaRadici}
	A.~Bacchetta, F.~Conti and M.~Radici,
  Phys.\ Rev.\  D {\bf 78}, 074010 (2008).  
    
\bibitem{Close:1988br}
  F.~E.~Close and A.~W.~Thomas,
  Phys.\ Lett.\  B {\bf 212}, 227 (1988).
  
\bibitem{Neubert}
  M.~Neubert and B.~Stech,
  Phys.\ Lett.\  B {\bf 231}, 477 (1989);
  Phys.\ Rev.\  D {\bf 44}, 775 (1991).
  
\bibitem{Bender:1996bb}
  A.~Bender, C.~D.~Roberts and L.~Von Smekal,
  Phys.\ Lett.\  B {\bf 380}, 7 (1996).
  
\bibitem{Klink:1998zz}
  W.~H.~Klink,
  Phys.\ Rev.\  C {\bf 58}, 3617 (1998);
  Phys.\ Rev.\  C {\bf 58}, 3587 (1998);
  R.~F.~Wagenbrunn, S.~Boffi, W.~Klink, W.~Plessas and M.~Radici,
  Phys.\ Lett.\  B {\bf 511}, 33 (2001);
  E.~P.~Biernat, W.~H.~Klink and W.~Schweiger,
  Few Body Syst.\  {\bf 49}, 149 (2011);
	W.~N.~Polyzou {\it et al.},
  Few Body Syst.\  {\bf 49}, 129 (2011).  
    
\bibitem{Lichtenberg:1979de}
  D.~B.~Lichtenberg and R.~J.~Johnson,
  Hadronic J.\  {\bf 2}, 1 (1979).
  
\bibitem{deCastro:1993sr}
  A.~S.~de Castro, H.~F.~de Carvalho and A.~C.~B.~Antunes,
  Z.\ Phys.\  C {\bf 57}, 315 (1993).  
            
\bibitem{Schafer:1993ra}
  T.~Sch\"{a}fer, E.~V.~Shuryak and J.~J.~M.~Verbaarschot,
  Nucl.\ Phys.\  B {\bf 412}, 143 (1994).
  
\bibitem{Cahill:1995ka}
  R.~T.~Cahill and S.~M.~Gunner,
  Phys.\ Lett.\  B {\bf 359}, 281 (1995).   
  
\bibitem{Lichtenberg:1996fi}
  D.~B.~Lichtenberg, R.~Roncaglia and E.~Predazzi,
  arXiv:hep-ph/9611428.
  
\bibitem{Burden:1996nh}
  C.~J.~Burden, L.~Qian, C.~D.~Roberts, P.~C.~Tandy and M.~J.~Thomson,
  Phys.\ Rev.\  C {\bf 55}, 2649 (1997).  
  
  
\bibitem{Maris}
  P.~Maris,
  Few Body Syst.\  {\bf 32}, 41 (2002) ;
  arXiv:nucl-th/0412059.
  
    	  
\bibitem{Orginos:2005vr}
  K.~Orginos,
  PoS {\bf LAT2005}, 054 (2006).            
                  
\bibitem{Flambaum:2005kc}
  V.~V.~Flambaum, A.~Holl, P.~Jaikumar, C.~D.~Roberts and S.~V.~Wright,
  Few Body Syst.\  {\bf 38}, 31 (2006).
  
\bibitem{Babich:2007ah} 
  R.~Babich, N.~Garron, C.~Hoelbling, J.~Howard, L.~Lellouch and C.~Rebbi,
  Phys.\ Rev.\  D {\bf 76}, 074021 (2007).
  
\bibitem{Eichmann:2008ef} 
  G.~Eichmann, I.~C.~Cloet, R.~Alkofer, A.~Krassnigg and C.~D.~Roberts,
  Phys.\ Rev.\  C {\bf 79}, 012202 (2009).
  
\bibitem{Lichtenberg:1981pp}
  D.~B.~Lichtenberg,
  Phys.\ Rev.\  {\bf 178}, 2197 (1969).

\bibitem{Godfrey:1985xj}
  S.~Godfrey and N.~Isgur,
  Phys.\ Rev.\  D {\bf 32}, 189 (1985).
  
\bibitem{DeSanctis:TBP}
  M.~De~Sanctis, J.~Ferretti, E.~Santopinto and A.~Vassallo,
  in preparation. 
  
\bibitem{BT}
	B.~Bakamjian and L.~H.~Thomas, 
	Phys. Rev. {\bf 92}, 1300 (1953). 
	
\bibitem{KP}
	B.~D.~Keister and W.~N.~Polyzou, 
	Adv. Nucl. Phys. {\bf 20}, 225 (1991).	
	
\bibitem{melde}
	T.~Melde, L.~Canton, W.~Plessas and R.~F.~Wagenbrunn, 
	Eur. Phys. J. {\bf A25}, 97 (2005).	 
  
\bibitem{Anisovich:2011fc} 
  A.~V.~Anisovich, R.~Beck, E.~Klempt, V.~A.~Nikonov, A.~V.~Sarantsev and U.~Thoma,
  Eur.\ Phys.\ J.\ A {\bf 48}, 15 (2012).       
  
\bibitem{Giannini:2001kb}
  M.~M.~Giannini, E.~Santopinto and A.~Vassallo,
  Eur.\ Phys.\ J.\  A {\bf 12} (2001) 447.
  
\bibitem{Glozman-Riska}
  L.~Y.~Glozman, D.~O.~Riska,
  Phys.\ Rept.\  {\bf 268} (1996) 263;
  L.~Y.~Glozman, W.~Plessas, K.~Varga, R.~F.~Wagenbrunn,
  Phys.\ Rev.\  D {\bf 58} (1998) 094030.  

\bibitem{Hecht:2002ej}
  M.~B.~Hecht, M.~Oettel, C.~D.~Roberts, S.~M.~Schmidt, P.~C.~Tandy and A.~W.~Thomas,
  Phys.\ Rev.\  C {\bf 65} (2002) 055204. 
  
\bibitem{Santopinto:2011mk}
  E.~Santopinto and G.~Galata,
  arXiv:1104.1518 [hep-ph].
  
\bibitem{Ebert}
  D.~Ebert, R.~N.~Faustov and V.~O.~Galkin,
  Phys.\ Rev.\  D {\bf 72}, 034026 (2005);
  Phys.\ Rev.\  D {\bf 84}, 014025 (2011).    
  
\bibitem{Chakrabarti:2010zz} 
  B.~Chakrabarti, A.~Bhattacharya, S.~Mani and A.~Sagari,
  Acta Phys.\ Polon.\ B {\bf 41}, 95 (2010).  
  
\bibitem{Roberts:2011cf}
  H.~L.~L.~Roberts, L.~Chang, I.~C.~Cloet and C.~D.~Roberts,
  Few Body Syst.\  {\bf 51}, 1 (2011).             
  
\end{thebibliography}
\end{document}